\documentclass[12pt]{iopart}

\begin{document}
\title{Non-relativistic proofs of the spin-statistics connection}

\author{Anil Shaji\dag\ and E. C. G. Sudarshan\ddag}

\address{\dag\ Center for Statistical Mechanics, Department of Physics, University of Texas, Austin, Texas 78712}

\address{\ddag\ Center for Particle Physics, Department of Physics, University of Texas, Austin, Texas 78712}

\ead{shaji@physics.utexas.edu}

\begin{abstract}
A recent paper by Peshkin \cite{peshkin} has drawn attention again to the problem of understanding the spin statistics connection in non-relativistic quantum mechanics. Allen and Mondragon \cite{allen} has pointed out correctly some of the flaws in Peshkin's arguments which are  based on the single valuedness under rotation of the wave functions of systems of identical particles. We examine carefully the claim that is made in the title of \cite{allen} that there can be ``no spin-statistics connection in non-relativistic quantum mechanics''. We show that we can derive the spin statistics connection for non-relativistic quantum field theories which have equations of motion of the Hamiltonian type based on $SU(2)$ invariance of the Lagrangian. The formalism and machinary of non-relativistic quantum field theory as opposed to usual quantum mechanics is necessary for constructing our proof. 
\end{abstract}

\submitto{\JPA}

\maketitle

\section{Introduction}
\label{intro}
Non-relativistic quantum systems like those in atomic physics, conduction electrons in metals, phonons, Cooper-pairs, quantum dots, Bose-Einstein condensates and many such others present the motivation for trying to understand the spin-statistics connection without having to use the full machinery of relativistic quantum field theory. The spectacular advances over the past decade in direct observations and manipulations of such systems involving a few non-relativistic quantum particles makes the need for such an understanding immediate. A variety of proposals based on the rotational properties and single valuedness of wave functions of many particle systems have been put forward \cite{broyles,bala,bacry,berry2} over the past couple of decades with this end in mind; Peshkin's \cite{peshkin} being the most recent. In response to some of these proposals (Peshkin's in particular), Allen and Mondragon has made the rather strong claim that there is ``No spin-statistics connection in non-relativistic quantum mechanics'' \cite{allen}. While they have correctly pointed out the problems with Peshkin's arguments we believe that it is important to elucidate clearly the precise role (if any) of relativity in establishing the spin-statistics connection \cite{duck2,duck1,ecg1,ecg2}. Our conclusion is that in the case of non-relativistic quantum fields in {\em three} space dimensions for which we can write down first order linear differential equations of the Hamiltonian form, the spin-statistics connection can be deduced without resorting to relativistic arguments. The cases of spin-$0$, spin-$\frac{1}{2}$ and spin-$1$ particles in three dimensions being the most significant and interesting ones because of their relevance to the non-relativistic quantum systems mentioned above. The other spins also obey the Michel mnemonic $P_{12}= (-1)^{2s}$ under permutation of particles.

The general line of argument used by many authors  to show the spin-statistics connection non-relativistically is to start from the assumption that the exchange of two quantum particles is equivalent to a physical rotation. This, along with certain specific requirements on the transformation properties of two particle wave functions under rotation, is claimed to lead to the required connection between spin and statistics. The pitfalls in such arguments, including those due to Peshkin, have already been discussed extensively in \cite{allen,duck2} and we shall not dwell on it any further. We present here a streamlined version of the arguments put forward by Sudarshan and Duck \cite{duck2,duck1} which establishes the spin-statistics connection, at least in a few important cases, without involving relativity or relativistic field theories or having to invoke rotational properties of many particle wave functions.  

\section{The key requirements for constructing the proof}
\label{requirements}

The proof of the spin-statistics connection that is presented in this paper utilizes the technique of second quantization and field theory in an essential manner as a means of dealing with many body quantum systems. Care must therefore be taken to distinguish the appearance of fields and field operators in the proof from the requirements of Lorentz invariance and relativistic causality. The fields that are introduced in the proof could perfectly well be non-relativistic like the Phonon field without impacting any of the arguments. The limitations that are introduced by retaining non-relativistic fields will be pointed out separately. We use quantum fields in our discussion because we want to view the exchange of particles that is so essential for the existence of the spin-statistics connection in very general terms. With the introduction of field operators that can create and annihilate particles at any point in space (and time) we can view the exchange of particles in a quantum system as destruction of particles at some points and their recreation elsewhere. We carefully and deliberately avoid the elaborate devices based on rotation operators and adiabatic transport used by some authors to obtain the spin-statistics connection and we use the formalism of non-relativistic field theory to construct our proof. The motivation for doing so being the belief that the spin-statistics connection must arise from the properties of the physical system independent of the {\em processes} that it goes through. In other words, the {\em kinematics} of the system should dictate the spin-statistics connection and not the dynamics. 

The original version of the proof which appeared in \cite{duck2} was in a form that could be applied to the relativistic case as well as the non-relativistic case. This introduced references to Lorentz invariance and the Lorentz group in the paper at places where they were not really essential. In this paper we present a non-relativistic version of the arguments and explore how far we can go with such a development.   

In what follows it is  assumed that we are dealing with a standard version of quantum field theory which is derivable from a Principle of Least Action based on a Lagrangian. It must be noted here that a slight generalization of the Principle of Least Action can lead to para-Bose and para-Fermi statistics in addition to the normal boson and fermion statistics. Such a generalization is only very briefly discussed in this paper so as to concentrate on the usual Fermi-Dirac and Bose-Einstein statistics and their origins. The formulation of the principle of least action in a form that is suitable for describing the dynamics of quantum systems and quantum fields will be discussed in section \ref{action}. 

Our arguments leading to the spin-statistics connection are confined at the outset to {\em three space dimensions}. The interesting possibilities that arise in lesser dimensions and the possible generality of an arbitrary number of dimensions are ignored in the interest of clarity. The three dimensional space admits multicomponent wave functions that have the usual transformation properties under rotations. The rotations that we consider are not restricted to the group of proper rotations in three dimensional space, $SO(3)$, but rather to its covering group $SU(2)$. The extension to $SU(2)$ is  needed for our discussion of the spin-statistics connection because we have to include both integral and half-integral spin representations of the rotation group and be able to treat them as proper representations. Since $SU(2)$ is a subgroup of the covering group of the Galilei group (as well as the Lorentz group), we are able to keep our arguments entirely non-relativistic. In fact, the transformation properties of the quantities that we are considering under $SU(2)$ turn out to be essential for the proof. 

 Along the lines of  Schwinger's proof \cite{schwinger} of the spin-statistics connection the following four conditions are imposed on the {\em kinematic part} of the Lagrangian density for each individual non-relativistic field appearing in the theories that we are considering.
\begin{enumerate}
\item The Lagrangian is {\em invariant under $SU(2)$ (and $SO(3)$) transformations} and corresponds to a field theory for fields, $\xi$, which are finite dimensional representations of the $SU(2)$ covering of the three dimensional rotation group. Since the $SU(2)$ is a subgroup of the covering of Galilei group, this requirement still holds if we let the fields be finite dimensional (irreducible) representations of the covering group pf the Galilei group. We have to consider the covering group rather than just the Galilei group in order to include spinors also as single valued representations. 
\item The Lagrangian is expressed in the {\em hermitian field basis}; $\xi = \xi^{\dagger}$.
\item It is at most linear in the {\em first time derivatives} of the field and the derivatives occur only in the kinematic term.
\item The {\em kinematic term is bilinear} in the field $\xi$
\end{enumerate} 

The last two requirements ensure that the Euler-Lagrange equations of motion derived from the Lagrangian are first order linear differential equations of the Hamiltonian form. The Dirac Lagrangian is already in the required form while for massive spin-$0$ and spin-$1$ fields satisfying the Klein-Gordon equation, the Duffin-Kemmer form of the equation is such a suitable one (see section \ref{kg}). A short discussion on how to convert a generic bilinear kinematic Lagrangian with higher order time derivatives in it into a form that is suitable for the present discussion is given in section \ref{linear}. 

The stipulation that the fields be in the Hermitian (real) representation introduces subtleties especially when dealing with internal charge-like (flavor) degrees of freedom of the field and when dealing with half integral spin fields. We will return to this issue in section \ref{kirchoff}. For the time being we consider the case where there is only one internal flavor index for the theory that we are considering. In what follows we focus on the kinematic parts of the Lagrangian written in a form that satisfies the four requirements laid out above. This is acceptable because we do not expect the spin-statistics connection to come out of interactions and other dynamical effects.

The generic (Schwinger) Lagrangian satisfying the four requirements has the form
\begin{eqnarray}
\label{eq:lagrangian}
{\mathcal{L}}& =& \frac{1}{2}K^0_{rs}(\xi_r \dot{\xi}_s- \dot{\xi}_r\xi_s) \nonumber \\
 & & + \;\; {\mbox{\em Terms containing space derivatives}} \nonumber \\
& & + \;\; {\mbox{\em Mass terms and nonlinear interaction terms}}
\end{eqnarray} 
The first term that is explicitly written down in the {\em kinematic} term on which we will focus our discussion. In the Lagrangian, $r$ and $s$ are spin indices that are summed over. The kinematic term is explicitly antisymmetrized with respect to the time derivative so that we avoid the possibility of the two terms together forming a total derivative. The terms containing the space derivatives need not be of this particular from and can contain higher derivatives and higher powers of the derivatives. The restrictions that are placed are exclusively on the time derivative terms which contain the kinematics of the system. A simple example is the Duffin-Kemmer form for spin-0 fields in which the Lagrangian is linear in the first time derivative but contains higher space derivatives apart from interaction terms (see section \ref{kg}).  No restrictions are placed on the mass term or on any interaction terms that may be present except to the extend that condition (iii) rules out velocity dependent potentials in the theories we are considering..  

The number of Hermitian fields that appear in the Lagrangian depend on the spin and possible charge of the physical field that we are considering. For a spin-$0$ field with no internal flavor index (charge) we need only one Hermitian field component, i.e. $r=s=1$. If we have a charged scalar field, then we need a pair of Hermitian fields with $r$, $s=1$ and an extra flavor index $\alpha =1 ,2$. The Lagrangian will then contain two pieces corresponding to each value of $\alpha$. For spin-$\frac{1}{2}$ and no charge, $r$, $s=1, \ldots 4$ as in the Majorana theory. The number of hermitian fields have to be doubled to eight, $r,s=1 \ldots 4$, $\alpha = 1,2$ when we are considering the Dirac theory for a spinor carrying a charge.  The Hermitian fields are used here (following Schwinger) expressly for the purpose of demonstrating the spin-statistics connection. For most other purposes this choice is unnecessarily cumbersome.

\subsection{Singular Lagrangians}
\label{singular}

If we require that the kinematic term of the Lagrangian be written in the form in (\ref{eq:lagrangian}) then we have to consider the possibility that $K^0_{rs}$ is a singular matrix. This would happen when some of the equations that we write down for the fields $\xi_r$ in the theory are not equations of motion (no time derivatives) but rather are equations of constraint. In the discussion that follows we will talk about the symmetry properties of $K^0_{rs}$ with respect to the indices $r$ and $s$. We will also talk about the commutation or anticommutation relation of the fields $\xi$ in terms of $K^0_{rs}$ and its inverse. Since we will be considering theories with primary constraints $K^0_{rs}$ will almost always be singular and because of this, {\em when we talk about the symmetry properties of $K^0_{rs}$ or about the inverse of $K^0_{rs}$ it must be understood that we are talking only about the non-singular part of the matrix}. The need for treating the non-singular part of the matrix separately is clear if we understand that in the form that we have written down the Lagrangian, there will in general be two kinds of fields in our theories. A set of fields $\xi^{(a)}$ for which we can define a {\em non-zero canonical conjugate} $\Pi^{(a)} \equiv \frac{\partial {\mathcal{L}}}{\partial \dot{\xi}^{(a)}}$  and another set of fields $\xi^{(b)}$ for which the canonical conjugate is identically zero. The first set, $\xi^{(a)}$, we call the canonical fields and the second set, the constraint variables. It is only the part of $K^0_{rs}$ that acts on the canonical field variables that we refer to when we talk about its inverse and its symmetry properties. The example of the neutral scalar field in section \ref{kg} will help to illustrate this point.

Apart from primary constraints that makes $K^0_{rs}$ singular there can also be secondary constraints. The presence of secondary constraints will make our discussion extremely complicated and so we will avoid it in this paper by assuming that there are {\em no secondary constraints} in the theories that we are dealing with in order to establish the spin-statistics connection. $K^0_{rs}$ could either be a real or complex Hermitian matirx depending on the fields in the theory we are considering.

\section{The crucial observation that leads to the spin-statistics connection}
\label{rotation}

The matrix $K^0_{rs}$ appearing in the Lagrangian in (\ref{eq:lagrangian}) is a numerical matrix that has no direct dependence on space and time coordinates. It depends on the spin labels only. It follows that if the Lagrangian has to be rotationally invariant (scalar) then the products of fields and their derivatives that appear in the Lagrangian are themselves scalars. It will turn out that $K^0_{rs}$ must have specific symmetry properties with respect to the spin indices $(r,s)$ in order for the Lagrangian to be nontrivial.

Obtaining the correct spin-statistics connection hinges on Sudarshan's observation that for the $SO(3)$ group of proper rotations (or rather its covering group $SU(2)$) in three dimensions, the representations belonging to {\em integral spin} have a bilinear scalar product that is {\em symmetric} in the indices of the factors. For example the scalar product of two real vectors is
\begin{equation}
  \label{spin1scalar}
  (V_1,V_2)= \sum_{j,k=1,2,3}V_{1j}V_{2k}\delta_{jk}.
\end{equation}
On the other hand, {\em half-integral spin} representations have {\em antisymmetric} bilinear scalar products. For instance, for the spin-$\frac{1}{2}$ case the scalar product of two Hermitian spinors is
\begin{equation}
  \label{spinhalfscalar}
  (\psi_1, \psi_2)= \sum_{r,s=1}^{4}\psi_{1r}\psi_{2s}(i \beta_{rs}),
\end{equation}
where 
\[ \beta_{rs} = \left( \begin{array}{cc} 0 & \hat{\sigma}_2 \\ \hat{\sigma}_2 & 0 \end{array} \right) \]
is an imaginary antisymmetric matrix corresponding the Majorana representation of the spinors. \footnote{for complex 2-component spinors the scalar product takes on the familar form $(\psi_1, \psi_2) =  \sum_{r,s=1,2} \psi_{1r} \psi_{2s}(i \sigma_y)_{rs}.$} . 

The requirement that the Lagrangian be $SU(2)$ invariant (in the Hermitian field basis) automatically requires that integral spin fields appear in the Lagrangian in symmetrized scalar combinations while half integral fields appear in antisymmetrized scalar combinations. These requirements put restrictions on the symmetry properties of the numerical matrices appearing in the Lagrangian. 

To investigate the symmetry requirements on $K^0_{rs}$ we focus our attention on the first (Kinematic) term of the Lagrangian in (\ref{eq:lagrangian}) i.e. 
\begin{equation}
  \label{eq:lagrangian2}
  {\mathcal{L}}_{kin} = \sum_{r,s}\frac{1}{2} K^0_{rs}( \xi_r \dot{\xi}_s - \dot{\xi}_r \xi_s).
\end{equation}
${\cal L}_{kin}$ may be rewritten in the following form
\[{\cal L}_{kin} = \frac{1}{2} \sum_{r,s} \xi_r \Lambda_{rs} \xi_s \quad ; \quad \Lambda_{rs} \equiv K^0_{rs} (\partial_t^{(r)} - \partial_t^{(s)}). \]
For tensor fields the scalar product constructed out of $\xi_r$ and $\xi_s$ that appears in the Lagrangian is symmetric in the indices $r$ and $s$. It follows that if the Lagrangian is to be non-trivial, $\Lambda_{rs}$ should also be symmetric in $r$ and $s$. 
\[ \Lambda_{rs} = \Lambda_{sr} \quad \Rightarrow  \quad K^0_{rs} (\partial_t^{(r)} - \partial_t^{(s)}) = - K^0_{sr}(\partial_t^{(r)} - \partial_t^{(s)}). \]
We see that because of the antisymmetry of the time derivative term, $K^0_{rs}$ has the opposite symmetry of $\Lambda_{rs}$ and so for tensor fields, $K^0_{rs}$ must be an antisymmetric matrix. 

For spinor fields the situation is reversed. Since the scalar product is antisymmetric, $\Lambda_{rs}$ also has to be antisymmetric for the Lagrangian to be non-trivial. Consequently for spinor fields, $K^0_{rs}$ must be a symmetric matrix.

Starting from the observation that for {\em integral spin} fields $K^0_{rs}$ must be antisymmetric in the spin indices $r$ and $s$ while it must be symmetric for {\em half integral spin} fields due to rotational invariance of  the Lagrangian; if we can now show by independent means that for {\em commuting fields} $K^0_{rs}$ must be antisymmetric and for {\em anticommuting fields} it must be symmetric we will obtain the proper spin-statistics connection. This is achieved using the Principle of Least Action for field quantities as formulated by Schwinger \cite{schwinger}.

\section{The principle of least action}
\label{action}

It was mentioned earlier that we are going to consider only a standard version of non-relativistic quantum field theory that is based on a suitable formulation of the principle of least action. In other words we deal with a canonically quantized theory that has a close correspondence with the Hamilton-Jacobi theory of classical systems. Since we are particularly interested in both commuting and anticommuting quantities, a suitably generalized dynamical principle based on the extended variation of the action integral is required; generalized, because after all, the classical version does not even admit anticommuting quantities.

Schwinger starts from the postulate that the generator of infinitesimal transformations on the eigenstates \{$| \zeta', \sigma \rangle$ \}  of a complete set of commuting operators of a quantized system/field is obtained by the extended variation of the quantities contained in the action integral
\begin{equation}
  \label{eq:action}
  I = \int_{\sigma_2}^{\sigma_1} (dx) \, {\mathcal{L}}[x].
\end{equation}
The integral is over an $n$-dimensional (space-time) domain with coordinates labeled by $x_k \, ; \, k = 1, \ldots, n$ and bounded by the $n-1$ dimensional surfaces $\sigma_1$ and $\sigma_2$. In the general case $\sigma_i$ are $(n-1)$ dimensional equal time slices in the $n$-dimensional space-time. As we already stated, in this discussion we consider {\em only} theories in three space and one time dimensions. Accordingly $\sigma_i$ are {\em three dimensional, equal time slices}. In equation (\ref{eq:action}), for ${\mathcal{L}}$ we use the specific form of the kinematic part of Lagrangian density that we wrote down in equation (\ref{eq:lagrangian}). 

The Principle of Stationary Action is then formulated for operator dynamical variables as the statement that the action integral operator is unaltered by infinitesimal variations of the field quantities in the interior of the region bounded by $\sigma_1$ and $\sigma_2$; it being dependent only on the variation of the complete commuting set of operators attached to the bounding surfaces.

The variation of the action integral can then be shown to be decomposable into two terms;
\begin{equation}
  \label{eq:actionvar}
  \delta I = \int_{\sigma_1}^{\sigma_2} dt d^3x \; {\mathcal{\delta L}}[x] + \int d^3x  \; ( {\mathcal{L}}[x, t_2] - {\mathcal{L}}[x, t_1]) \delta x.  
\end{equation}
The first term is an integral over the space-time domain ${\cal{D}}$ bounded by $\sigma_1$ and $\sigma_2$. The second term is a ``surface'' term integrated only over the three spaces at times $t_1$ and $t_2$. Setting the variation of the action in the interior of the space-time domain $\cal{D}$ to zero leads to the Euler-Lagrange equations of motion for the fields while the surface variation term is treated as the generator of infinitesimal transformations on the system. 

We need not go into the details of this development here but what is pertinent to our discussion is to note, as indeed pointed out by Schwinger, that $\delta {\mathcal{L}}$ should be treated carefully in the quantum case since the order of the operators that appear in ${\mathcal{L}}$ must not be altered in the course of effecting the variation. Accordingly, we recognize that the commutation properties of $\delta \xi_r$ that appear in the variation of the action based on the Lagrangian density in equation (\ref{eq:lagrangian}) are involved in the consequences of the extended variation of the action integral. In other words we must, at this point, make the explicit assumption that the commutation properties of $\delta \xi$ with respect to $\xi_r$ and the structure of the Lagrange function must be connected in a consistent fashion.

We now choose to simplify matters as much as possible and as before focus in on the kinematic part $ {\mathcal{L}}_{kin}$ of  our general Lagrangian density. We are interested in the present context only on the variation of the action that is brought about by the variation in the field quantities $\xi$ themselves. We will not therefore consider the variation of the action integral operator that is brought out by a change in the coordinates $x$ or time $t$. The variations that we consider here have the form 
\begin{eqnarray}
  \label{eq:variation}
  \xi_k \rightarrow \xi_k + \delta \xi_k \nonumber \\
  \dot{\xi}_k \rightarrow \dot{\xi}_k + \frac{d (\delta \xi_k)}{dt}.
\end{eqnarray}
The ensuing variation in the action is then given by
\begin{equation}
  \label{eq:var1}
  \delta I_{\delta \xi} = \frac{1}{2} \int dt \, d^3x \left( \frac{\partial {\mathcal{L}}}{\partial \xi_r} \delta \xi_r + \frac{\partial{\mathcal{L}}}{\partial \dot{\xi}_r}\delta \dot{\xi}_r \right).
\end{equation}
The integral is over a three dimensional space and one dimensional time domain. Also, the above equation must again be understood in a symbolic sense with the exact position of the variation $\delta \xi$ in each term to be fixed later in an appropriate fashion. Using the identity,
\[ \frac{\partial{\mathcal{L}}}{\partial \dot{\xi}_r}\delta \dot{\xi}_r = \frac{\partial \;}{\partial t} \left(\frac{\partial{\mathcal{L}}}{\partial \dot{\xi}_r}\delta \xi_r \right) -  \frac{\partial \;}{\partial t} \left(\frac{\partial{\mathcal{L}}}{\partial \dot{\xi}_r} \right) \delta \xi_r , \]
and using Gauss' theorem we rewrite the variation in the action integral as 
\begin{equation}
  \label{eq:var2}
   \delta I_{\delta \xi} = \frac{1}{2} \int dt \, d^3x \left\{  \frac{\partial {\mathcal{L}}}{\partial \xi_r} - \frac{\partial \;}{\partial t} \left(\frac{\partial{\mathcal{L}}}{\partial \dot{\xi}_r} \right) \right\} \delta \xi_r \nonumber  +  \frac{1}{2} \int_{\sigma} d^3x \frac{\partial{\mathcal{L}}}{\partial \dot{\xi}_r}\delta \xi_r.
\end{equation}
Setting the first term (variation in the interior of the domain ${\cal{D}}$ bounded by $\sigma$) to zero we obtain the Euler-Lagrange equations for the hermitian fields  $\xi_r$. It is the surface variation terms that are of interest to us as the generators of infinitesimal transformations on the field quantities themselves. Note that since we are considering non-relativistic quantities, the surface integral is over an equal-time slice with volume element $d^3x$.
For the specific form for ${\mathcal{L}}_{kin}$ that we are considering in equation (\ref{eq:lagrangian2}), the (surface) variation term is given by
\begin{equation}
  \label{eq:var3}
   \delta I_{\delta \xi} = \frac{1}{2} \int_{\sigma} d^3x \sum_{r,s} K^0_{rs}(\xi_r \delta \xi_s - \delta \xi_r \xi_s). 
\end{equation}
Since $ \delta I_{\delta \xi}$ is the generator of infinitesimal transformations of the field quantities $\xi$ attached to the surface $\sigma$  we require
\begin{equation}
  \label{eq:comm1}
  [ \xi_n,  \delta I_{\delta \xi}] = i \hbar \delta \xi_n
\end{equation}
or
\begin{equation}
  \label{eq:comm2}
  \frac{1}{2} \int_\sigma d^3x \left[ \xi_n ,  \sum_{r,s} K^0_{rs}(\xi_r \delta \xi_s - \delta \xi_r \xi_s ) \right] = i \hbar \delta \xi_n.
\end{equation}
Expanding out the commutator we obtain
\begin{equation}
  \label{eq:comm3}
   \frac{1}{2} \int_\sigma d^3x \sum_{r,s}K^0_{rs} ( \xi_n  \xi_r \delta \xi_s - \xi_n \delta \xi_r \xi_s -  \xi_r \delta \xi_s  \xi_n +  \delta \xi_r \xi_s \xi_n) =  i \hbar \delta \xi_n.
\end{equation}
Now we are in a position to assume that the fields appearing in equation (\ref{eq:comm3}) are either commuting fields or anticommuting fields and investigate what restrictions (if any) each assumption places on the matrix $K^0_{rs}$. The multiplication of field operators defined at the same point in the above equations can lead to divergence problems due to the non-zero vacuum expectation value of the field modes. A discussion of such problems is beyond the scope of this paper but we point out that subtracting out the vacuum expectation value does not affect the commutation relations of the fields and therefore it does not affect the development presented in the sections that follow.  

\subsection{Commuting fields} \label{commute}

Let us assume now that the fields $\xi$ are bosonic fields and we consider variations in them. The significant step is that we assume $\delta \xi$ {\em commutes} with everything. Using this assumption we can rewrite the left hand side of equation (\ref{eq:comm3}) as 
\begin{equation}
  \label{eq:comm4}
   \frac{1}{2} \int_\sigma d^3x \sum_{r,s}K^0_{rs} \{ ( \xi_n  \xi_r \delta \xi_s - \xi_r \xi_n \delta \xi_s)  - (\xi_n \xi_s \delta \xi_r -  \xi_s \xi_n \delta \xi_r) \} 
\end{equation}
Putting in the space coordinates ${\bf x}$ and ${\bf y}$ of the fields explicitly we can rewrite the previous expression in terms of commutation relations of the fields as 
\begin{equation} 
\label{comm5a}
 \frac{1}{2} \int_\sigma d^3x \sum_{r,s}K^0_{rs} \{ [ \xi_n({\bf y}), \xi_r({\bf x})] \delta \xi_s({\bf x})   - [\xi_n({\bf y}), \xi_s({\bf x})] \delta \xi_r({\bf x}) \}.
\end{equation}
Exchanging the indices $i$ and $j$ in the second term we can rewrite the expression in equation (\ref{comm5a}) as
\begin{equation}
\label{eq:comm5}
[\xi_n, \delta I_{\delta \xi}] =  \int_\sigma d^3x \sum_{s} \delta \xi_s({\bf x}) \left[ \xi_n({\bf y}) , \,\frac{1}{2} \sum_r (K^0_{rs} - K^0_{sr})\xi_r({\bf x}) \right]  
\end{equation}

From equation (\ref{eq:comm5}) we see explicitly that $[\xi_n, \delta I_{\delta \xi}] = 0$ if $K^{0}_{ij}$ is symmetric. So to satisfy the condition that the surface variation of the action is the generator of infinitesimal transformations we need $K^{0}_{rs}$ to be an {\em antisymmetric} matrix if the fields $\xi_r$ (and $\delta \xi_r$) are bosonic commuting fields. The possibility that $K^0_{rs}$ be neither symmetric nor antisymmetric ($K^0_{rs} \ne \pm K^0_{sr}$) in the spin indices is already excluded by the rotational invariance of the Lagrangian density that we require.

We can verify the consistency of the discussion presented above by noting that 
\[ \Pi_l \equiv \frac{\partial {\mathcal{L}}}{\partial \dot{\xi}_l} = \frac{1}{2} \sum_{rs}K^0_{rs}(\xi_r \delta_{sl} - \delta_{rl}\xi_s) =  \frac{1}{2}\sum_r(K^0_{rl}-K^0_{lr})\xi_r \]
Therefore equation (\ref{eq:comm5}) becomes
\begin{equation}
  \label{eq:comm5b}
  [\xi_n, \delta I_{\delta \xi}] =  \int_\sigma d^3x \sum_{s} \delta \xi_s({\bf x}) [\xi_n({\bf y}), \Pi_j({\bf x})] = i\hbar \delta \xi_n ({\bf y}) .
\end{equation}
This leads to the usual canonical commutation relations for the (canonical, not constraint) fields $\xi_r$;
\[ [\xi_n({\bf y}), \Pi_j({\bf x})] =  i \hbar \delta^{(3)}({\bf x} - {\bf y}) \delta^{(K)}_{nj} \]
where $\delta^{(K)}_{nj}$ denotes the Kronecker delta function and $\delta^{(n)}$ denotes the Dirac delta function. Note that since we have used only hermitian fields with a kinematic term in the Lagrangian that is linear in the time derivatives we see that the canonical momentum conjugate to the field variable $\xi_r$ is also a linear function of $\xi_r$. 

\subsection{Anticommuting fields} \label{anticommute}
If we assume fermionic anticommutation relations  instead of commutation relations for the fields $\xi$ and assume that $\delta \xi$ {\em anticommutes} with everything, the analogue of equation (\ref{eq:comm4}) that we obtain from (\ref{eq:comm3}) is 
\begin{equation}
  \label{eq:comm6}
    \frac{1}{2} \int_\sigma d^3x \sum_{r,s}K^0_{rs} \{ ( \xi_n  \xi_r \delta \xi_s + \xi_r \xi_n \delta \xi_s)  + (\xi_n \xi_s \delta \xi_r +  \xi_s \xi_n \delta \xi_r) \} 
\end{equation}
This can be simplified as before to 
\begin{equation}
  \label{eq:comm7}
   [\xi_n , \delta I_{\delta \xi}] = \int_\sigma d^3x \sum_{s} \delta \xi_s({\bf x}) \left\{ \xi_n({\bf y}), \frac{1}{2} \sum_r (K^0_{rs} + K^0_{sr})\xi_r({\bf x}) \right\}  
\end{equation}
In this case we see that $ [\xi_n , \delta I_{\delta \xi}] = 0$ if $K^0_{rs}$ is antisymmetric. Thus for anticommuting fermionic fields we require $K^0_{rs}$ to be a {\em symmetric} matrix to be consistent with the dynamical principle.

\section{The spin-statistics connection}

We have seen that the rotational invariance of the kinematic part of the Lagrangian density requires that the matrix $K^0_{rs}$ be antisymmetric for integral spin fields while it be symmetric for half-integral fields. The Schwinger Principle of Least Action on the other hand requires that $K^0_{rs}$ be antisymmetric for commuting (bosonic) fields and symmetric for anticommuting (fermionic) fields. This leads us to the conclusion that {\em fields with integral spin must be bosonic while fields with half-integral spin must be fermionic}. This is the spin-statistics connection. 

\section{Limits of applicability of the proof}

The discussion of a non-relativistic derivation of the spin-statistics connection that we have presented is based on a few requirements or postulates. These postulates, which were listed earlier, and we delineate the limits that they place on the applicability of the derivation to various situations. We investigate a few such issues in this section

\subsection{Internal flavor indices and Kirchoff's Principle} \label{kirchoff}
The arguments presented in the previous three sections leading to the spin-statistics connection rely on the symmetry, or antisymmetry, that is required independently by $SU(2)$ invariance of the Lagrangian and by the Principle of Least Action, of the numerical matrix $K^0_{rs}$. We now investigate the possibility of changing the symmetry requirements on $K^0_{rs}$ with respect to $r$ and $s$ that is demanded by rotational invariance through the introduction of internal symmetry indices (flavors) on the fields $\xi_r$. In other words we relax the condition that the kinematic part of the Lagrangian be diagonal in flavor indices that was assumed in the previous sections. We consider a Lagrangian with the following form for the kinematic term
\begin{equation}
  \label{eq:lagrangianc}
  {\mathcal{L}}_{kin} = \frac{1}{2} \sum_{r,s,\alpha,\beta}K^0_{\alpha r, \beta s} (\xi_r^{\alpha} \dot{\xi}_s^{\beta} - \dot{\xi}_r^{\alpha} \xi_s^{\beta}). 
\end{equation}
$\alpha$ and $\beta$ denoting the values of an internal charge-like degree of freedom (flavor), $Q_{\alpha}$. Let us restrict $\alpha = \beta = 1,2$; this being the simplest interesting case. Writing out all the terms in $ {\mathcal{L}}_{kin}$ we obtain,
\begin{eqnarray}
  \label{eq:lagrangiand}
   {\mathcal{L}}_{kin} &=& \frac{1}{2} \sum_{rs} K^0_{1r,1s}(\xi_r^1 \dot{\xi}_s^1 - \dot{\xi}_r^1 \xi_s^1) + \frac{1}{2} \sum_{rs} K^0_{2r,2s}(\xi_r^2 \dot{\xi}_s^2 - \dot{\xi}_r^2 \xi_s^2) \nonumber \\
& & + \frac{1}{2} \sum_{rs} K^0_{1r,2s}(\xi_r^1 \dot{\xi}_s^2 - \dot{\xi}_r^1 \xi_s^2) +  \frac{1}{2} \sum_{rs} K^0_{2r,1s}(\xi_r^2 \dot{\xi}_s^1 - \dot{\xi}_r^2 \xi_s^1). 
\end{eqnarray}
In the construction of our proof, the symmetry properties of the scalar products of fields of the form $\xi_r \dot{\xi}_s$ that appear in $ {\mathcal{L}}_{kin}$ fixed the symmetry properties of $K^0_{rs}$ with respect to the spin indices $r$ and $s$ on which the rotation ($SU(2)$) group elements act. With the introduction of the extra labels $\alpha$ and $\beta$ we look for the possibility of inverting the symmetry requirements on $K^0_{rs}$ with respect to $r$ and $s$  by antisymmetrizing $ {\mathcal{L}}_{kin}$ with respect to the new labels. If ${\mathcal{L}}_{kin}$ is symmetric with respect to $\alpha$ and $\beta$ our previous arguments go through without any change and this is not interesting to us here.

We therefore choose $K^0_{rs}$ to be antisymmetric in $\alpha$, $\beta$, i.e:
\begin{equation}
  \label{eq:kalphabeta}
  K_{1r,2s}= -K_{2r,1s} \quad ; \quad K_{1r,1s}=K_{2r,2s}=0.
\end{equation}
Equation (\ref{eq:lagrangiand}) then reduces to
\begin{equation}
  \label{eq:lagrangiane}
   {\mathcal{L}}_{kin}=\frac{1}{2} \sum_{rs} K_{1r,2s} ( \xi^1_r \dot{\xi}_s^2 - \dot{\xi}_r^1 \xi_s^2 - \xi_r^2 \dot{\xi}_s^1 + \dot{\xi}_r^2 \xi_s^1). 
\end{equation}
The term within parentheses in (\ref{eq:lagrangiane}) is symmetric under the interchange of $r$ and $s$ if each scalar product appearing in it is symmetric in $r$ and $s$. This means that if $\xi_r$ are tensor fields then $K^0_{1r,2s}$ must be {\em symmetric} in the indices $r$ and $s$ to ensure the rotational invariance of ${\mathcal{L}}_{kin}$. On the other hand the Principle of Least Action would lead to anticommutation relations for $\xi_r$ in order to be consistent with the form of $K^0_{1r, 2s}$  which is now symmetric with respect to the indices $r$ and $s$.  So one could come to the conclusion that $\xi_r$ must be anticommuting fields of integral spin!

A similar analysis will reveal that if the scalar products appearing in the term within parentheses in (\ref{eq:lagrangiane}) are antisymmetric in $r$ and $s$ then $K^0_{1r,2s}$ must be {\em antisymmetric} in order to make ${\mathcal{L}}_{kin}$ rotationally invariant. This, along with the Principle of Least Action will lead one to the conclusion that $\xi_r$ must be commuting spinor fields. 

We now show that {\em antisymmetrization} of the Lagrangian on such internal indices leads to states with negative norm in the theories that we are considering. Stipulating that such negative norm states (negative Hilbert space metric!) cannot be present in any physical theory will eliminate the possibility of obtaining an inverted spin-statistics connection as described above. 

\subsubsection{Negative norm states}

The Lagrangian in (\ref{eq:lagrangiane}) can be written in matrix form as
\begin{equation}
  \label{eq:lagrangianf} 
\fl  {\mathcal{L}}_{kin} = \frac{1}{2} (\xi_r^1 \, , \, \xi_s^2) \left( \begin{array}{cc} 0 & K^0_{rs} \\ -K^0_{rs} & 0 \end{array} \right) \left( \begin{array}{c} \dot{\xi}_r^1 \\ \dot{\xi}_s^2 \end{array} \right)  -  \frac{1}{2} (\dot{\xi}_r^1 \, , \, \dot{\xi}_s^2) \left( \begin{array}{cc} 0 & K^0_{rs} \\ -K^0_{rs} & 0 \end{array} \right) \left( \begin{array}{c} \xi_r^1 \\ \xi_s^2 \end{array} \right),
\end{equation}
with the matrices appearing on the right hand side assumed to be in block form, $r$ and $s$ running over however many species of fields $\xi$ we may have. We can write ${\mathcal{L}}_{kin}$ in a more compact form as
\begin{equation}
{\mathcal{L}}_{kin} = (\xi_r^1 \, , \, \xi_s^2) \Lambda^{12}(\stackrel{\rightarrow}{\partial_t} - \stackrel{\leftarrow}{\partial_t})\left( \begin{array}{c} \xi_r^1 \\ \xi_s^2 \end{array} \right) \quad ; \quad \Lambda^{12} = \left( \begin{array}{cc} 0 & K^0_{rs} \\ -K^0_{rs} & 0 \end{array} \right).
\end{equation}
$\Lambda^{12}$ is an antisymmetric matrix with {\em zero trace}. We can always find a unitary transformation $S$ that diagonalizes $\Lambda^{12}$ (See \ref{appendixa} for an example). Using the transformation $S$, the Lagrangian ${\mathcal{L}}_{kin}$ can be rewritten in a form that is diagonal in the flavor index as follows,
\begin{eqnarray}
\label{eq:lagrangiandiag}
{\mathcal{L}}_{kin}& = & (\xi_r^1 \, , \, \xi_s^2)SS^{-1} \Lambda^{12}(\stackrel{\rightarrow}{\partial_t} - \stackrel{\leftarrow}{\partial_t})SS^{-1}\left( \begin{array}{c} \xi_r^1 \\ \xi_s^2 \end{array} \right) \nonumber \\
& = &(\tilde{\xi}_r^1 \, , \, \tilde{\xi}_s^2)D^{12}(\stackrel{\rightarrow}{\partial} - \stackrel{\leftarrow}{\partial})\left( \begin{array}{c} \tilde{\xi}_r^1 \\ \tilde{\xi}_s^2 \end{array} \right)
\end{eqnarray}
where $D^{12}$ is now a diagonal matrix of the form 
\begin{equation}
\label{d}
D^{12} = \left( \begin{array}{cc} \mu \tilde{K}_{rs}^0 & 0 \\ 0 & -\mu \tilde{K}^0_{rs} \end{array} \right). 
\end{equation}
and
\begin{equation}
\label{eq:tildexi}
(\xi^1_r \, , \, \xi^2_s)S = (\tilde{\xi}^1_r \, , \, \tilde{\xi}^2_s).
\end{equation}
The matrix $D^{12}$ has two eigenvalues $\pm \mu$ of equal magnitudes but of opposite sign since $\Lambda^{12}$ was traceless. In terms of the transformed fields $\tilde{\xi}_r$ the Lagrangian now has the form
\begin{equation}
\label{eq:lagrangiang}
{\mathcal{L}}_{kin} = \frac{1}{2} \mu \sum_{r,s} \tilde{K}^0_{rs} ( \tilde{\xi}^1_r \dot{\tilde{\xi}}^1_s - \dot{\tilde{\xi}}^1_r \tilde{\xi}^1_s) - \frac{1}{2} \mu \sum_{r,s} \tilde{K}^0_{rs} ( \tilde{\xi}^2_r \dot{\tilde{\xi}}^2_s - \dot{\tilde{\xi}}^2_r \tilde{\xi}^2_s)
\end{equation} 
What is important to note in the above equation is that the two terms corresponding to $\alpha=1$ and $\alpha=2$ come with opposite signs. The surface variation of the action integral computed using the form the Lagrangian that is diagonal in the flavor index corresponding to the variation of the fields $\tilde{\xi}^{\alpha}_r$ is
\begin{equation}
\delta \tilde{I}_{\delta \tilde{\xi}} = \frac{1}{2} \mu \int_{\sigma} d^3 x  \sum_{r,s} \tilde{K}^0_{rs}[ (\tilde{\xi}^1_r \delta \tilde{\xi}^1_s - \delta \tilde{\xi}^1_r \tilde{\xi}^1_s) - ( \tilde{\xi}^2_r \delta \tilde{\xi}^2_s - \delta \tilde{\xi}^2_r \tilde{\xi}^2_s)].
\end{equation}
To satisfy the basic commutation relation 
\[ [ \tilde{\xi}^{\alpha}_n \, , \,\delta \tilde{I}_{\delta \tilde{\xi}}] = i \hbar \delta \tilde{\xi}^{\alpha}_n, \]  
the symmetry of $K^0_{rs}$ leads to canonical {\em anticommutation} relations of the form 
\begin{eqnarray}
\label{eq:cancomm}
\{ \tilde{\xi}^1_n({\bf x}) \, , \, \frac{1}{2}\sum_r (K^0_{rm}+K^0_{mr})\tilde{\xi}^1_r ({\bf y}) \} = i \hbar \delta^{(K)}_{nm} \delta^{(3)}({\bf x} - {\bf y}) \; \, \nonumber \\
\{ \tilde{\xi}^2_n({\bf x}) \, , \, \frac{1}{2}\sum_r (K^0_{mr}+K^0_{rm})\tilde{\xi}^2_r ({\bf y}) \} = -i \hbar \delta^{(K)}_{nm} \delta^{(3)}({\bf x} - {\bf y}).
\end{eqnarray}
If we now write 
\begin{equation}
\label{field1}
\tilde{\xi}^1_r = \tilde{\xi}^1_r(-) + \tilde{\xi}^1_r(+) = \sum_k \frac{1}{\sqrt{2 \omega_{k,r}}} ( a_{k,r} e^{-i(kx-\omega t)} + b_{k,r}^{\dagger} e^{i(kx-\omega t)} ) 
\end{equation}
and 
\begin{equation}
\label{field2}
\tilde{\xi}^2_r = \tilde{\xi}^2_r(-) + \tilde{\xi}^2_r(+) = \sum_k \frac{1}{\sqrt{2 \omega_{k,r}}} ( c_{k,r} e^{-i(kx-\omega t)} + d_{k,r}^{\dagger} e^{i(kx-\omega t)} ),
\end{equation}
the anticommutation relations that follow from (\ref{eq:cancomm}) for the creation and annihilation operators which appear in $\tilde{\xi}^1_r$ and $\tilde{\xi}^2_r$ are of the form
\begin{equation}
\label{eq:acomm1}
\{ a, b^{\dagger} \} = \{ b^{\dagger}, a \} \sim 1 
\end{equation}
and
\begin{equation}
\label{eq:bcomm1}
\{ c, d^{\dagger} \} = \{ d^{\dagger}, c \} \sim -1. 
\end{equation}
Taking the vacuum expectation value of equation (\ref{eq:bcomm1}) one sees that the Hilbert space of $\tilde{\xi}^2_r$ contains states with negative norm, i.e.  $\langle 1^2_r | 1^2_r \rangle = -1$. 

What is important to note here is that diagonalization of any matrix $\Lambda^{\alpha \beta}$ which is antisymmetric in charge like degrees of freedom $Q_{\alpha_i}$ and which has zero trace will lead to Lagrangians with kinematic terms having both positive and negative signs. The terms with negative signs will correspond to fields with Hilbert spaces that contains negative norm states. The source of the negative norm states can therefore be traced to the antisymmetrization with respect to the flavors $Q_{\alpha}$. Since we cannot have states with negative norm in our theory we can exclude the possibility of antisymmetrizing ${\mathcal{L}}_{kin}$ with respect to flavor indices and thereby inverting the spin-statistics connection.    

Because we are dealing with non-relativistic quantum fields and operators there is a way of avoiding the negative norm states even while antisymmetrizing on internal indices. This can be done by considering the creation and annihilation fields $\tilde{\xi}_r^{\alpha}(\pm)$ in isolation as the basic field operators in our theory instead of combining them into a single field. For relativistic fields, Lorentz invariance and the requirement of locality ensure that all physical fields come as combinations of creation and annihilation parts. For non-relativistic fields however there is no such requirement {\em a priori}. Instead, we have to restrict ourselves to Hermitian fields. This automatically realizes the {\em Kirchoff's Principle} which makes sure that even non-relativistic fields appear as combinations of creation and annihilation parts.

\subsubsection{Kirchoff's Principle}
\label{Kirchoff}
When we study the thermodynamics of radiation in equilibrium, we assume that there is a continual emission and absorption of the radiation by matter. Furthermore we know that the emissivity and absorptivity of matter are proportional to each other: this is Kirchoff's Principle. In terms of ordinary quantum mechanics what this means is that when we couple harmonic oscillator degrees of freedom, the coupling is in terms of the coordinate $q \sim (a + a^{\dagger})/2$ rather than coupling the creation and annihilation operators separately and independently. In an analogous manner we realize Kirchoff principle for the non-relativistic fields by. The requirement that the probabilities of creating and destroying the quanta of the fields that we are considering are proportional to each other ensures that the creation and annihilation fields do not appear independently in the Lagrangians that we are considering. Kirchoff's principle thus plugs a potential loophole in our derivation of the spin-statistics connection. Kirchoff's Principle is also a consequence of our use of hermitian fields since to construct a real field we need {\em both the creation and annihilation parts to appear together.}  

As far as relativistic fields are concerned we may deduce Kirchoff's principle fome the requirement of locality of the fields. This automatically makes sure that the creation and annihilation fields always come together in any physical field that appears in a relativistic field theory. Splitting a field into the creation and annihilation parts is a nonlocal operation in relativistic theory. But in both non-relativistic and relativistic field theory, Hermitian fields assure compliance with Kirchoff's principle

\subsection{Nature of the variation $\delta \xi_r$}
\label{deltaxi}

In our discussion of the Principle of Least Action for quantum mechanical operators and fields in section \ref{action}, we considered the variation of the action when the fields $\xi_r$ are varied by $\delta \xi_r$. To get commutation or anticommutation relations for the fields and to obtain the correct spin-statistics connection we made the choice that $\delta \xi_r$ commutes with every quantity that appears in the variation of the action when $\xi_r$ denote commuting fields and that $\delta \xi_r$ anticommutes with everything if $\xi_r$ are anticommuting quantities. This is required if we want to keep the varied fields $\xi^{\prime}_r = \xi_r + \delta \xi_r$ such that they have {\em exactly the same commutation or anticommutation properties as $\xi_r$}.  

Choosing the arbitrary variation $\delta \xi_r$ to be a quantity that commutes with everything is easy enough because all one has to do is to make sure that $\delta \xi$ is a c-number. On the other hand, choosing $\delta \xi_r$ such that it anticommutes with all the field quantities that appear in $\delta I_{\delta \xi}$ is not so straightforward. For an even number of anticommuting fields, $\xi_r$, one possible choice to make $\delta \xi_r$ proportional to the product of all the fields, i.e
\[ \delta \xi_r \sim \epsilon_r \, \Pi_{i=1}^{2n} \xi_i \quad ; \quad r =1,2, \ldots 2n. \]
$\delta \xi_r$ will then anticommute with all the quantities that appear in the variation of the action \cite{cartan}. 

Restricting $\delta \xi_r$ to be either commuting or anticommuting with everything is essential for the canonical quantization of commuting or anticommuting fields \cite{schweber,weinberg1}. But it is possible to make the choice of $\delta \xi_r$ more general. If we consider the fields $\xi_r$ defined on the equal time slice $\sigma$ as forming a complete set of operators on $\sigma$, a natural generalization of $\delta \xi_r$ is to choose it as a linear combination of $\xi_r$,
\begin{equation}
  \label{eq:parastat1}
  \delta \xi_r({\bf x}) = \sum_s \varepsilon_{rs} \xi_s ({\bf x})  
\end{equation}
where $\varepsilon_{rs}$ are infinitesimal $c$-number coefficients. The basic commutation relation in (\ref{eq:comm2}) can be rewritten as
\begin{equation}
  \label{eq:comm2a}
  \frac{1}{2} \int_\sigma d^3x \left[ \xi_n ,  \sum_{i,j,k} K^0_{rs}(\xi_i \varepsilon_{jk}\xi_k - \varepsilon_{ik} \xi_k \xi_j ) \right] = i \hbar \sum_{k}\varepsilon_{nk} \xi_k.
\end{equation}
Without working out the details we can immediately see that eq. (\ref{eq:comm2a}) will lead to {\em trilinear} commutation or anticommutation relations among $\xi_r$. These generalized commutation and anticommutation relations, due to Wigner and Green \cite{wigner,green}, lead to para-Bose and para-Fermi systems. A discussion of the para-Bose and para-Fermi fields is beyond the scope of this paper; we will return to it in a separate publication.

\subsection{Lagrangians linear in the first time derivative}
\label{linear}
The requirement that we placed on the kinematic terms of the Lagrangian that it be linear in the {\em first time derivative} of the fields and bilinear in the fields themselves might seem to be too restrictive at first. It can be shown that any bilinear kinetic Lagrangian containing terms with higher order time derivatives can be reduced to the form we require even if such a form might not be the most convenient or elegant one. 

The original construction due to Ostrogradsky \cite{ostro,whittaker,pais} can be sketched using the notation used in the present paper as follows. We start with a system with an equation of motion for a single field $\phi$  that contains higher order time  derivatives of the form
\begin{equation}
  \label{eq:higher}
  {\cal{F}}(\partial_t)\phi = 0 \quad ; \quad \partial_t = \frac{\partial \; }{\partial t}
\end{equation}
where ${\cal{F}}(\partial_t)$ is a polynomial of {\em finite} degree. If we restrict ourselves to reversible motions, ${\cal{F}}$ is an even function of its argument. We can be more general and consider ${\cal{F}}$ to be a polynomial in both space and time derivatives but since the discussion in this paper focuses mostly on the kinematic terms of the Lagrangian, we shall not be concerned with this possibility. The equation of motion (\ref{eq:higher}) may be derived from a Lagrangian density,
\begin{equation}
  \label{eq:lagrangianhi}
  {\mathcal{L}}_{kin} \sim -\phi {\cal{F}}(\partial_t) \phi.
\end{equation}
Making ${\mathcal{L}}_{kin}$ linear in the first order time derivatives is achieved by introducing a set of auxiliary fields defined by
\begin{equation}
  \label{eq:aux1}
  \xi_i = \partial_t^{n-1} \phi \quad ; \quad n = 1, \ldots, N.
\end{equation}
where $\partial_t^N$ is the highest derivative of $\phi$ appearing in the Lagrangian. We can in fact go further and define the canonical conjugates of $\xi_i$ through the linear combinations:
\begin{equation}
  \label{eq:aux2}
  \Pi_{\xi_i} = \frac{\delta {\mathcal{L}}_{kin}}{\delta (\partial^i_t \phi)},
\end{equation}
where
\[ \frac{\delta  {\mathcal{L}}}{\delta x} \equiv \frac{\partial  {\mathcal{L}}}{\partial x} - \partial \frac{\partial  {\mathcal{L}}}{\partial (\partial_t x)} + \partial^2 \frac{\partial  {\mathcal{L}}}{\partial (\partial_t^2 x)} - \ldots \]
Using the newly defined field variables one can define the hamiltonian ${\cal{H}}$ for the system will then contain all the non time derivative terms of ${\mathcal{L}}$. 

Starting from ${\cal{H}}$ one can then obtain linear differential equations of the Hamiltonian type for $\xi_r$. In the present context all we are really interested in is in writing ${\mathcal{L}}_{kin}$ in the form that we require in section \ref{requirements} using the newly defined auxiliary fields $\xi_r$. The simple example of a neutral scalar field $\phi$ satisfying the ($2^{nd}$ order) Schr\"{o}dinger-Klein-Gordon equation illustrates how this can be done.

\subsubsection{Example: Schr\"{o}dinger-Klein-Gordon equation for the neutral scalar field}
\label{kg}
The Schr\"{o}dinger-Klein-Gordon equation for a free neutral scalar field is a second order differential equation, 
\begin{equation}
  \label{eq:kg}
  (\partial_{\mu}\partial^{\mu} - m^2) \phi = 0.
\end{equation}
The corresponding Lagrangian,
\begin{equation}
  \label{eq:kgl}
  {\mathcal{L}}= \frac{1}{2}(\dot{\phi}^2 - (\nabla \phi)^2) + \frac{1}{2}m^2\phi^2 
\end{equation}
is quadratic in $\dot{\phi}$ and is not of the form we require. Duffin and Kemmer \cite{duffin,kemmer} showed that we can rewrite the Lagrangian as 
\begin{equation}
  \label{eq:kgl2}
  {\mathcal{L}} = \bar{\psi}(i \beta_{\mu} \partial^{\mu} - m ) \psi 
\end{equation}
where 
\[ \psi = (\phi,  \dot{\phi} \, , \,  \partial_x \phi \, , \,  \partial_y \phi \, , \, \partial_z \phi)^T \quad ; \quad \bar{\psi} = (\phi,  \dot{\phi} \, , \,  - \partial_x \phi \, , \, -\partial_y \phi \, , \,  - \partial_x \phi). \]
and the matrices $\beta_{\mu}$ satisfy the trilinear relations
\begin{eqnarray*}
\beta_{\mu}^3 & =& \beta_\mu , \\
\beta+{\mu}\beta_{\nu}\beta_{\mu} & = & \beta_{\mu} \quad ; \quad \mu \neq \nu, \\
\beta_{\mu}\beta_{\nu}^2 + \beta_{\nu}^2\beta_{\mu} & = & \beta_{\mu} \quad ; \quad  \mu \ne \nu, \\
\beta_{\mu}\beta_{\nu}\beta_{\lambda} + \beta_{\lambda} \beta_{\nu} \beta_{\mu} & =& 0 \quad ; \quad \mu \ne \nu \ne \lambda.
\end{eqnarray*} 
We note here in passing that the components of $\psi$ are all Hermitian and a possible specific choice of the $\beta$ matrices has
\[ \beta_0 = \left( \begin{array}{ccccc} \; 0 \; & \; i \;  & \; 0 \; & \; 0 \; & \; 0 \; \\ -i & 0 & 0 & 0 & 0 \\ 0 & 0 & 0 & 0 & 0 \\ 0 & 0 & 0 & 0 & 0 \\  0 & 0 & 0 & 0 & 0 \end{array} \right) \]
which is an antisymmetric matrix as we require for integral spin fields. All the four $\beta$-matrices are listed in \ref{appendixb}.

Note that $\beta_0$ is a singular matrix with no inverse since the last three rows and columns in it contain only zeros. But then the equations that we write down for the new fields $\partial_x\phi$, $\partial_y \phi$ and $\partial_z \phi$ do not contain any time derivatives and hence are {\em not equations of motion}. In other words, $\partial_x \phi$, $\partial_y \phi$ and $\partial_z \phi$ can be classified as constraint variables while $\phi$ and $\dot{\phi}$ are the canonical field variables (one being the conjugate of the other in this case). Accordingly when we talk about the symmetry properties of $\beta_0$ and its inverse we are referring to the $2 \times 2$ block
\[ \tilde{\beta}_0 = \left( \begin{array}{cc} 0 & i \\ -i & 0 \end{array} \right). \]

\section{Conclusion}
\label{concl}
We have shown in this paper that for the general class of (not necessarily non-relativistic) fields which can be canonically quantized, there exists a proof of the spin-statistics connection that does not depend on relativistic arguments. All that is required is the $SU(2)$ invariance of the Lagrangian. To avoid a possible way of inverting the spin-statistics connection obtained by our arguments, we have to explicitly exclude systems with negative norm states. Both requirements are reasonable and justified in our point of view especially when we are dealing with non-relativistic systems. We have also touched upon the possibility of how para-Bose and para-Fermi systems can be included within the scope of our discussions.

\ack
The authors thank Dr. Tom Jordan for pointing out the need for a clear elucidation of a non-relativistic derivation of the spin-statistics connection and initiating the work presented here. One of the authors (Anil Shaji) also thanks Mr. C. Krishnan for lively and illuminating discussions. We also thank the referees for pointing out the errors and omissions in our arguments in the original version of the paper.

\appendix
\section{Diagonalizing a Lagrangian that is antisymmetrized on internal indices.}
\label{appendixa}
Starting from the Lagrangian in equation (\ref{eq:lagrangiane}) the simplest case we can consider (with $K^0_{1r,2s}$ symmetric) is to set $r=s=1$. In matrix form this Lagrangian looks like
\begin{equation}
\label{eq:appendixa1}
{\mathcal{L}}_{kin} = (\xi^1_1 \, , \, \xi^2_1) \left( \begin{array}{cc} 0 & 1 \\ -1 & 0 \end{array} \right) (\stackrel{\rightarrow}{\partial_t} - \stackrel{\leftarrow}{\partial_t}) \left( \begin{array}{c} \xi_1^1 \\ \xi_1^2  \end{array} \right).
\end{equation}
with $K^0_{11,21}$ set to $1$. The matrix appearing in (\ref{eq:appendixa1}) is diagonalized by the transformation 
\begin{equation}
\label{eq:appendixa2}
\frac{1}{\sqrt{2}} \left( \begin{array}{cc} 1 & i \\ 1 & -i \end{array} \right) \left( \begin{array}{cc} 0 & 1 \\ -1 & 0 \end{array} \right) \left( \begin{array}{cc} 1 & 1 \\ -i & i \end{array} \right) \frac{1}{\sqrt{2}} =   \left( \begin{array}{cc} -i & 0 \\ 0 & i \end{array} \right).
\end{equation}
The corresponding transformations on the fields are
\begin{eqnarray}
\label{eq:appendixa3}
\tilde{\xi}^1 = \frac{\xi^1 - i \xi^2}{\sqrt{2}} \\
\tilde{\xi}^2 = \frac{\xi^1 + i \xi^2}{\sqrt{2}}
\end{eqnarray}
(dropping the redundant space-time subscript `1'). 

In terms of the transformed fields, the Lagrangian becomes
\begin{equation}
\label{eq:appendixa4}
{\mathcal{L}}_{kin} = \frac{1}{2}[(\tilde{\xi}^1 \dot{\tilde{\xi}}^1 - \dot{\tilde{\xi}}^1 \tilde{\xi}^1) - (\tilde{\xi}^2 \dot{\tilde{\xi}}^2 - \dot{\tilde{\xi}}^2 \tilde{\xi}^2)]. 
\end{equation} 
In equation (\ref{eq:appendixa4}) we see explicitly how antisymmetrizing on the internal index leads to a (flavor diagonalized) Lagrangian that has {\em kinematic} terms with negative signs which leads to negative norm states. Note here that $\tilde{\xi}^1$ and $\tilde{\xi}^2$ are not hermitian fields.

\section{The $\beta$ matrices in the Duffin-Kemmer equation}
\label{appendixb}
\begin{eqnarray*}
\beta_0 =  \left( \begin{array}{ccccc} \; 0 \; & \; i \;  & \; 0 \; & \; 0 \; & \; 0 \; \\ -i & 0 & 0 & 0 & 0 \\ 0 & 0 & 0 & 0 & 0 \\ 0 & 0 & 0 & 0 & 0 \\  0 & 0 & 0 & 0 & 0 \end{array} \right) \quad ; \quad & \beta_1 =  \left( \begin{array}{ccccc} \; 0 \; & \; 0 \;  &  -i   & \; 0 \; & \; 0 \; \\ 0 & 0 & 0 & 0 & 0 \\ -i & 0 & 0 & 0 & 0 \\ 0 & 0 & 0 & 0 & 0 \\  0 & 0 & 0 & 0 & 0 \end{array} \right) \\
\beta_2 =  \left( \begin{array}{ccccc} \; 0 \; & \; 0 \;  & \; 0 \; & -i  & \; 0 \; \\ 0 & 0 & 0 & 0 & 0 \\ 0 & 0 & 0 & 0 & 0 \\ -i & 0 & 0 & 0 & 0 \\  0 & 0 & 0 & 0 & 0 \end{array} \right) \quad ; \quad &  \beta_3 =  \left( \begin{array}{ccccc} \; 0 \; & \; 0 \;  & \; 0 \; & \; 0 \; & -i \\ 0 & 0 & 0 & 0 & 0 \\ 0 & 0 & 0 & 0 & 0 \\ 0 & 0 & 0 & 0 & 0 \\  -i & 0 & 0 & 0 & 0 \end{array} \right)
\end{eqnarray*}

\end{document}